\documentclass[11pt,epsf,amssymb]{article} \textheight =23 truecm
\def\lsim{\raise0.3ex\hbox{$<$\kern-0.75em\raise-1.1ex\hbox{$\sim$}}}
\def\gsim{\raise0.3ex\hbox{$>$\kern-0.75em\raise-1.1ex\hbox{$\sim$}}}
\def\noi{\noindent}  \def\bea{\begin{eqnarray}}
\def\eea{\end{eqnarray}} \def\beq{\begin{equation}}
\def\eeq{\end{equation}} 
\def\beeq{\begin{eqnarray}} \def\eeeq{\end{eqnarray}} \def\R{ {\rm R
\kern -.31cm I \kern .15cm}} \def\C{ {\rm C \kern -.15cm \vrule
width.5pt \kern .12cm}} \def\Z{ {\rm Z \kern -.27cm \angle \kern
.02cm}} \def\N{ {\rm N \kern -.26cm \vrule width.4pt \kern .10cm}}
\def\1{{\rm 1\mskip-4.5mu l} }

\usepackage{graphics} \usepackage{graphicx} \usepackage{epsfig}

\title{  Analysis of data on $B$ decays into two light vector mesons \protect\\}

\author{R. Aleksan$^1$, L. Oliver$^2$ \\
\footnotesize $^1$IRFU, CEA, Universit\'e Paris-Saclay, 91191 Gif-sur-Yvette cedex, France \\
\footnotesize $^2$IJCLab, P\^ole Th\'eorie, CNRS/IN2P3 et Universit\'e Paris-Saclay, \\
\footnotesize B\^at. 210, 91405 0rsay, France }

\begin{document}
\maketitle

\begin{abstract}

An important experimental effort has been accomplished in recent years in the measurement of rates, polarization and CP observables in $B$ decays into two light vector mesons. On the theoretical side, after a very consistent effort done within the framework of QCD Factorization, the comparison of the theory with the present experimental data has not been updated, to our knowledge. In the present paper we compare this wealth of data to the theory, in tree color-allowed, tree color-suppressed and pure penguin decays, and present predictions for the observables that have not yet been measured in these decays. In our fits we find acceptable values of $\chi^2$ for the branching ratios and for the direct CP asymmetries. However, this is not the case for the longitudinal polarization fractions, essentially due to disagreement between theory and experiment for the modes $B_{d,s}\to K^{*0}\overline{K}^{*0}$ and $B_d \to \rho^+ \rho^-$. Although we rely on previous work by other authors, we summarize the formalism so that the paper is self-contained and its results can be checked.

\end{abstract}

\section{{\Large Introduction}}

The physics of $B$ meson decays into two vector mesons is very rich due to the polarization degrees of freedom. In this paper we focus more specifically on the decays $B_{d,s}$ into two light vector mesons that have been experimentally studied at BaBar, Belle and LHCb~\cite{PDG-2022}. Within the Standard Model (SM), the QCD Factorization (QCDF) scheme offers a good framework to describe these decays. We therefore propose to study to which extent QCDF is able to reproduce the data.\par
\noi In Section 2 we summarize the branching fractions, polarization fractions and CP asymmetries for these $B$ decays into two light vectors mesons.\par
\noi In Section 3 we descride the main lines of the QCD Factorization scheme of $B$ decays into two light vector mesons as studied by \cite{KAGAN,BENEKE-2007,CHENG-YANG,BUCHALLA}.\par
\noi In Section 4 we formulate the decay amplitudes in terms of QCDF coefficients.\par
\noi In Section 5 we give the results of QCDF for the branching ratios, polarization fractions and CP asymmetries and compare with the existing data. \par

\noi In the Appendices we give useful formulas on the Wilson expansion (Appendix A), on the QCDF coefficients $a_i^h$ (Appendix B) and on the QCDF annihilation coefficients $b_i^h$ (Appendix C) for both helicities $h = 0, -$.\par
\vskip 10pt
\noi Although our calculations are based on well-established previous works \cite{KAGAN,BENEKE-2007,CHENG-YANG,BUCHALLA}, we describe our formalism in some detail in the Appendices in order to facilitate possible future verifications of our final numerical results by other interested authors.

\section{Branching ratios, polarization fractions and CP asymmetries for $\overline{B}_{d,s}$ decays into two light vector mesons}

To be definite, let us write down the helicity amplitudes
\beq
\label{1e} 
\overline{A}_L = A[\overline{B} \to V_1(0) V_2(0)]\ , \qquad \qquad \overline{A}_\pm = A[\overline{B} \to V_1(\pm) V_2(\pm)]
\eeq

\noi where $V_1$ and $V_2$ are respectively the emitted and the produced vector mesons.\par 

\noi Then the transversity amplitudes read
\beq
\label{2e} 
\overline{A}_\parallel = {1 \over \sqrt{2}} (\overline{A}_+ + \overline{A}_-) \ , \qquad \qquad \overline{A}_\perp = {1 \over \sqrt{2}} (\overline{A}_+ - \overline{A}_-)
\eeq

\noi with the corresponding transversity rate fractions $f_L , f_\parallel$ and $f_\perp$, satisfying 
\beq
\label{2-1e} 
f_L + f_\parallel + f_\perp = 1
\eeq

\noi Since the final quark is dominantly left-handed because of the $V-A$ structure of the Standard Model (SM), heavy quark symmetry implies the hierarchy
\beq
\label{2-2e} 
\overline{A}_L : \overline{A}_- : \overline{A}_+ = 1 : {\Lambda_{QCD} \over m_b} : \left({\Lambda_{QCD} 
\over m_b}\right)^2
\eeq

\vskip 3 truemm

\noi As underlined in detail by Beneke et al. \cite{BENEKE-2007}, the transverse amplitude $\overline{A}_-$ is suppressed by a factor $m_{V_2}/m_B$ relative to $\overline{A}_L$, and the axial and vector contributions to $\overline{A}_+$ cancel in the heavy quark limit, implying the hierarchy (\ref{2-2e}).\par

\noi The limit
\beq
\label{3e} 
\overline{A}_+ = 0
\eeq

\noi implies \cite{BENEKE-2007},
\beq
\label{4e} 
f_\parallel = f_\perp
\eeq

\noi The hierarchy (\ref{2-2e}) points out to 
\beq
\label{5e} 
f_\parallel \simeq f_\perp << f_L
\eeq

\noi Present data on branching ratios and polarization fractions are given in Table~\ref{tab:B_decays}.

\vskip 3 truemm
\begin{table}[htb]
\centering
\footnotesize

\begin{tabular}{|c|c|c|c|c|c|}

\hline 
Decay mode & $\rm{BR}\ (\times 10^{-6})$ & $f_L$ & $f_\parallel$ & $f_\perp$ & $A_{CP}$\\ \hline \hline 
$\overline{B}_d \to \omega \omega$ & $1.2 \pm 0.4$ &
 & & & \\ \hline
$\overline{B}_d \to \overline{K}^{*0} \omega$ & $2.0 \pm 0.5$ &
$0.69 \pm 0.11$ & $ $ & $ 0.10 \pm 0.13$ & $ 0.45 \pm 0.25$ \\ \hline
$\overline{B}_d \to \overline{K}^{*0} K^{*0}$ & $0.83 \pm 0.24$ &
$0.74 \pm 0.05$ & $ $ & $ $ & $ $ \\ \hline
$\overline{B}_d \to \overline{K}^{*0} \phi$ & $10.0 \pm 0.5$ &
$0.497 \pm 0.017$ & $ $ & $0.224 \pm 0.015$ & $0.0 \pm 0.04$ \\ \hline
$\overline{B}_d \to \overline{K}^{*0} \rho^0$ & $3.9 \pm 1.3$ &
$0.173 \pm 0.026$ & $ $ & $0.40 \pm 0.04$ & $-0.06 \pm 0.09$ \\ \hline
$\overline{B}_d \to \overline{K}^{*-} \rho^+$ & $10.3 \pm 2.6$ &
$0.38 \pm 0.13$ & $ $ & $ $ & $0.21 \pm 0.15$ \\ \hline
$\overline{B}_d \to \rho^- \rho^+$ & $27.7 \pm 1.9$ &
$0.990 \pm 0.020$ & $ $ & $ $ & $ $ \\ \hline
$\overline{B}_d \to \rho^0 \rho^0$ & $0.97 \pm 0.24$ &
$0.71 \pm 0.09$ & $ $ & $ $ & $ $ \\ \hline \hline
$\overline{B}_s \to \phi \rho^0$ & $0.27 \pm 0.08$ &
 & &  & \\ \hline
$\overline{B}_s \to \phi \phi$ & $18.5 \pm 1.7$ &
$0.38 \pm 0.01$ & $ $ & $ 0.29 \pm 0.01$ & $ $ \\ \hline
$\overline{B}_s \to \overline{K}^{*0} K^{*0}$ & $11.10 \pm 2.67$ &
$0.24 \pm 0.04$ & $ 0.30 \pm 0.05 $ & $ 0.38 \pm 0.12$ & $ $ \\ \hline
$\overline{B}_s \to K^{*0} \phi$ & $1.14 \pm 0.30$ &
$0.51 \pm 0.04$ & $0.21 \pm 0.05$ & $ $ & $ $ \\ \hline \hline
$B^- \to K^{*-} \omega$ & $ < 7.4$ &
$0.41 \pm 0.19$ &  &  & $0.29 \pm 0.35$ \\ \hline
$B^- \to K^{*-} K^{*0}$ & $0.91 \pm 0.29$ &
$0.82 \pm 0.20$ &  & $ $ & $ $  \\ \hline
$B^- \to K^{*-} \phi$ & $10.0 \pm 2.0$ &
$0.50 \pm 0.05$ &  & $0.20 \pm 0.05$ & $-0.01 \pm 0.08$ \\ \hline
$B^- \to K^{*-} \rho^0$ & $4.6 \pm 1.1$ &
$0.78 \pm 0.12$ &  & $ $ & $0.31 \pm 0.13$  \\ \hline
$B^- \to \overline{K}^{*0} \rho^-$ & $9.2 \pm 1.5$ &
$0.48 \pm 0.08$ &  & $ $ & $-0.01 \pm 0.16$  \\ \hline
$B^- \to \rho^0 \rho^-$ & $24.0 \pm 1.9$ &
$0.95 \pm 0.02$ &  & $ $ & $-0.05 \pm 0.05$  \\ \hline
$B^- \to \omega \rho^-$ & $15.9 \pm 2.1$ &
$0.90 \pm 0.06$ &  & $ $ & $-0.20 \pm 0.09$  \\ \hline

\end{tabular}
\caption{\label{tab:B_decays} Data on the rates, polarization fractions and CP asymmetries for $\overline{B}_{d,s,u}$ decays into two light vectors mesons~\cite{PDG-2022}.}

\end{table}

\noi It is important to underline that the data for penguin-dominated $B$ decays are in conflict with the expected hierarchy (\ref{2-2e}) if one uses Naive Factorization, that predicts a large longitudinal fraction $f_L \sim 1$, in consistency with the data for tree decay modes, e.g. $\overline{B}_d \to \rho \rho$.

\section{QCD Factorization scheme for $B$ decays into two light vector mesons}

To go beyond the Naive Factorization calculation, the natural theoretical scheme is QCD Factorization (QCDF), that starts from short distance operators with their Wilson coefficients, including the Penguin and EW Penguin operators, given in Appendix A. Next, this scheme includes NLO $O(\alpha_s)$ Vertex corrections $V$, Penguin diagrams $P$, Hard Spectator diagrams $H$ and Weak Annihilation $A$, see Fig. 1 for the example $\overline{B}_s \to \phi \phi$.\par

\includegraphics[scale=0.55]{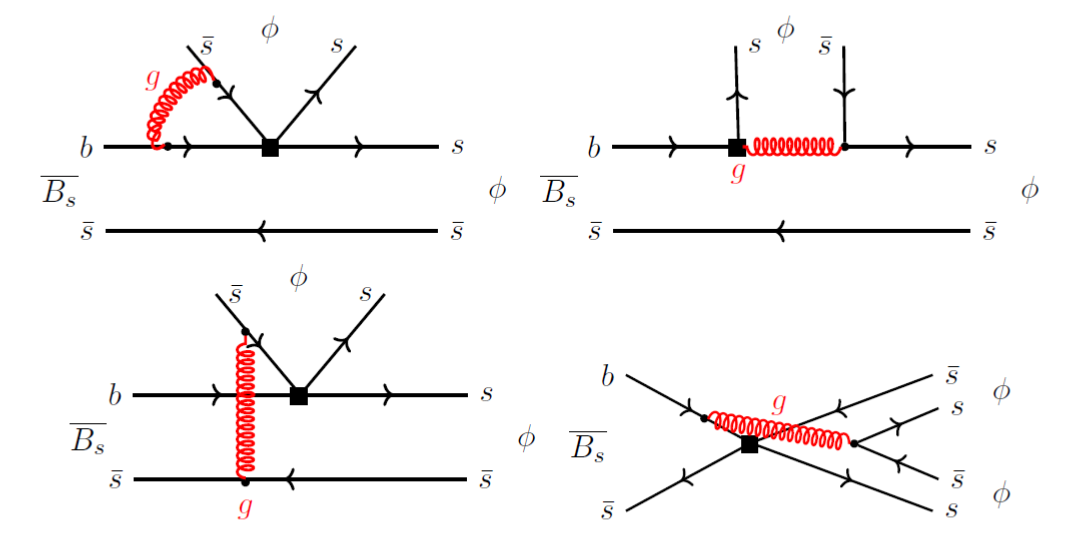}

Fig. 1. Typical NLO $\alpha_s$ corrections of the QCDF scheme. From left to right and from above to below : corrections of the types Vertex $V$, Penguin $P$, Hard Spectator $H$ and Annihilation $A$.

\vskip 3truemm

\noi The QCDF scheme includes NLO non-factorizable corrections of $O(\alpha_s)$, and the corresponding {\it effective} Wilson coefficients $a_i^h$ are {\it helicity-dependent}. One could then expect that QCDF could have consequences on the pattern of the polarization data of Table~\ref{tab:B_decays}.
This is indeed the case, as has been shown by the pionnering papers of Kagan \cite{KAGAN}, Beneke et al. \cite{BENEKE-2007} and Cheng and Yang \cite{CHENG-YANG}.\par

\noi The Wilson coeffficients $C_i\ (i=1,... 10)$ at NLO \cite{BBNS-01}, for $\mu = m_b$, with $m_b(m_b) = 4.2\ \rm{GeV}$, $\Lambda_{\overline{MS}}^{(5)} = 0.225\ \rm{GeV}$, are given in~\ref{tab:11} of Appendix A.\par

\noi Schematically, the matrix elements have the structure \cite{BBNS-01,BBNS-99,BBNS-00},
$$<V_1 V_2\mid O_i \mid \overline{B} >\ = \left[F^{\overline{B} \to V_1} T_i^I \otimes f_{V_2} \Phi_{V_2} + (V_1 \leftrightarrow V_2) \right] $$
\beq
\label{6e}
+\ T_i^{II} \otimes f_B \Phi_B \otimes f_{V_1} \Phi_{V_1} \otimes f_{V_2} \Phi_{V_2}
\eeq

\noi The first term depends on the form factor $F^{B \to V_1}$ and decay constant $f_{V_2}$, and the second is the Annihilation, dependent on the three decay constants $f_B, f_{V_1}, f_{V_2}$.\par 

\noi In QCD Factorization, the coefficients $a_i^h$ depend on the helicity. From formulas (11) of \cite{KAGAN}, (53) of \cite{BENEKE-2007} and (24) of \cite{BUCHALLA} the coefficients $a_i^{p,h} (p = u, c)$ of QCD Factorization write, {\it in a simplified way},
$$a_i^{p,h}(V_1 V_2) = \left(C_i(\mu) + {C_{i\pm 1}(\mu) \over N_c} \right) N_i^h(V_2) + {C_{i\pm 1}(\mu) \over N_c} {C_F \alpha_s(\mu) \over {4 \pi}}\ V_i^h(V_2)$$
\beq
\label{7e} 
+ {C_{i\pm 1}(\mu_h) \over N_c} {C_F \alpha_s(\mu_h) \over {4 \pi}} {4 \pi^2 \over N_c}\ H_i^h(V_1V_2) + P_i^{p,h}(V_2)
\eeq

\vskip 4 truemm

\noi where $h$ is the helicity, $C_i$ are the short distance Wison coefficients tabulated in Appendix A, the upper (lower) signs correspond to $i$ odd (even), $N_i$ are normalization factors, $V_i$ are the vertex corrections, $H_i$ are the hard scattering corrections, and $P_i^p$ are the Penguin corrections with $p = u, c$. In this expression $V_1$ and $V_2$ are respectively the produced and the emitted meson and we have made explicit that the scale is not the same in front of the Hard scattering term $H_i^h$, as explained in Appendix A.\par 
\noi The final expressions of $a_i^{p,h}(V_1 V_2)$ are given in Appendix B, that contain some extra contributions that slightly modify the expression (\ref{7e}) separately for the helicities $h = 0$ and $h = -$. The coupling $\alpha_s$ and the Wilson coefficients for both scales $\mu$ and $\mu_h$ are given respectively by (\ref{C-7-3e}) and Tables~\ref{tab:11} and~\ref{tab:12}.
In the limit of disregarding the NLO $\alpha_s$ corrections, the coefficients (\ref{7e}) become the usual combinations of short distance Wilson coefficients, e.g.  $a_4^{p,h} \to C_4 + {C_3 \over N_c}$\ ,... \par

\noi The annihilation coefficients $b_i^h$ are given in terms of a few building blocks by the expressions
$$b_3^h = {C_F \over N_c^2} \left[C_3 A_1^{ih} + C_5 (A_3^{ih} + A_3^{fh}) +N_c C_6 A_3^{fh} \right] \ \ \ \ $$
\beq
\label{8e}
b_4^h = {C_F \over N_c^2} \left[C_4 A_1^{ih} + C_6 A_2^{ih} \right] \qquad \qquad \qquad \qquad \ \ \ \ \ 
\eeq
$$b_3^{EWh} = {C_F \over N_c^2} \left[C_9 A_1^{ih} + C_7 (A_3^{ih} + A_3^{fh}) +N_c C_8 A_3^{fh} \right]$$
$$b_4^{EWh} = {C_F \over N_c^2} \left[C_{10} A_1^{ih} + C_8 A_2^{ih} \right] \qquad \qquad \qquad \qquad $$

\noi where $A_i^{ih}$ and $A_i^{fh}$ are given in Appendix C for $h = 0$ and $h = -$.

\vskip 3truemm

\noi The decay amplitudes for $\Delta S = 0$ two-body transitions read, with a sum over the quarks $p = u, c$,
\beq
\label{9-1e}
A(\overline{B} \to V_1 V_2, h) = \sum_{p = u, c} \lambda_p \left(S^{p,h}\ A^h_{V_1 V_2} + T^{p,h}\ B_{V_1 V_2}^h \right)
\eeq

\noi while the decay amplitudes for $\Delta S = 1$ two-body read
\beq
\label{9-2e}
A(\overline{B} \to V_1 V_2, h) = \sum_{p = u, c} \lambda'_p \left(S^{p,h}\ A^h_{V_1 V_2} + T^{p,h}\ B_{V_1 V_2}^h \right)
\eeq

\noi where $\overline{B} = \overline{B}_d, \overline{B}_s$ or $B^-$. The first term proportional to $A^h_{V_1 V_2}$ corresponds to the direct contribution, while the second term proportional to $B^h_{V_1 V_2}$ corresponds to the annihilation contribution. The CKM coefficients are given by
\beq
\label{10e}
\lambda_p = V_{pb}V^*_{pd}\ , \qquad \qquad \lambda'_p = V_{pb}V^*_{ps} \qquad \qquad (p = u, c)
\eeq

\noi and numerically in Table~\ref{tab:2}.

\vskip 6 truemm
\begin{table}
\centering
\footnotesize
\begin{tabular}{|c|c|c|}
\hline 
$\lambda_u$ & $V_{ub} V^*_{ud}$ & $(0.001286 \pm 0.000152) - (0.003268 \pm 0.000115) i$ \\ \hline
$\lambda_c$ & $V_{cb} V^*_{cd}$ & $(-0.009174 \pm 0.000161) + (5.370 \pm 0.267) \cdot 10^{-6} i$ \\ \hline
$\lambda'_u$ & $V_{ub} V^*_{us}$ & $(0.000299 \pm 0.000035) - (0.000760 \pm 0.000027) i$ \\ \hline
$\lambda'_c$ & $V_{cb} V^*_{cs}$ & $(0.03944 \pm 0.00069) + (1.249 \pm 0.006) \cdot 10^{-6} i$ \\ \hline

\end{tabular}

\caption{\label{tab:2}  CKM parameters.}

\end{table} 

\noi To get the CP conjugated amplitudes $A(B \to V_1 V_2, h)$ one needs to make the replacements $\lambda_p \to \lambda_p^*, \lambda'_p \to \lambda_p^{\prime*} (p = u, c)$ in (\ref{9-1e},\ref{9-2e}).\par 

\noi In eqns. (\ref{9-1e},\ref{9-2e}), $S^{p,h}$ are linear combinations of the direct coefficients (\ref{7e}), while $T^{p,h}$ are linear combinations of the annihilation coefficients (\ref{8e}). Both quantities must be specified for each decay mode.

\noi The coefficients in (\ref{9-1e},\ref{9-2e}) for $h = 0$ of the direct diagram and the annihilation diagram are given by
\beq
\label{11-1e}
A^0_{V_1V_2} = i {G_F \over \sqrt{2}}\ m_{B}^2 A_0^{B \to V_1} (m_{V_2}^2) f_{V_2} \ , \qquad \qquad B^0_{V_1V_2} = i {G_F \over \sqrt{2}} f_{B} f_{V_1} f_{V_2}
\eeq

\noi while the coefficients for the helicity $h = -$ of the direct diagram and the annihilation diagram are given by
\beq
\label{11-2e}
A^-_{V_1V_2} = i {G_F \over \sqrt{2}}\ m_{B} m_{V_1} F_-^{B \to V_1} (m_{V_2}^2) f_{V_2} \ , \qquad \qquad B^-_{V_1V_2} = i {G_F \over \sqrt{2}} f_{B} f_{V_1} f_{V_2}
\eeq

\noi where the masses, decay constants and form factors are given in Tables~\ref{tab:3} and~\ref{tab:4}.

\vskip 5 truemm

\begin{table}
\centering
\footnotesize

\begin{tabular}{|c|c|c|c|}
\hline 
Particle $x$ & $m_x$ (MeV) & $f_x$ (MeV) & $\tau_x$ (s) \\ \hline
$\overline{B}_u$ & $5279.34 \pm 0$ & $190. \pm 5$ & $1.638 \times 10^{-12} \pm0$ \\ \hline
$\overline{B}_d$ & $5279.65 \pm 0$ & $190. \pm 5$ & $1.519 \times 10^{-12} \pm 0$ \\ \hline
$\overline{B}_s$ & $5366.88 \pm 0$ & $130. \pm 5$ & $1.515 \times 10^{-12} \pm 0$ \\ \hline

\end{tabular}

\caption{\label{tab:3} B meson parameters.}

\end{table}

\begin{table}
\centering
\footnotesize
\begin{tabular}{|c|c|c|c|c|c|c|}
\hline 
Particle $x$ & $m_x$ (MeV) & $f_x$ (MeV) & $A_0^{B_d \to x}$ & $F_-^{B_d \to x}$ & $A_0^{B_s \to x}$ & $F_-^{B_s \to x}$ \\ \hline
$\rho^0$ & $775.\pm 0$ & $209 \pm 5$ & $0.30 \pm 0.05$ & $0.55 \pm 0.06$ & n/a & n/a \\ \hline
$\omega$ & $782.\pm 0$ & $187 \pm 5$ & $0.25 \pm 0.05$ & $0.50 \pm 0.06$ & n/a & n/a \\ \hline
$K^{*0}$ & $895.\pm 0$ & $218 \pm 5$ & $0.39 \pm 0.05$ & $0.68 \pm 0.06$ & $0.33 \pm 0.05$ & $0.53 \pm 0.06$ \\ \hline
$\phi$ & $1019.5\pm 0$ & $221 \pm 5$ & n/a & n/a & $0.38 \pm 0.05$ & $0.65 \pm 0.06$ \\ \hline

\end{tabular}

\caption{\label{tab:4} Light meson parameters and heavy-to-light form factors for $q^2 = 0$.}

\end{table}

\section{The decay amplitudes in terms of the QCDF coefficients $a_i^{p,h}, b_i^h$}

Our aim is to compare the QCDF predictions with decay modes of Table~\ref{tab:B_decays}, for which there are experimental data for at least the branching ratio and the longitudinal polarization fraction. To this aim we need the expressions of the amplitudes of these modes in terms of the coefficients $a_i^{p,h}$ and $b_i^h$. These expressions are given very explicitely by the excellent work of Bartsch et al. \cite{BUCHALLA} in the formulas (49)-(76) for the part of the amplitudes that depend on the direct decay coefficients $a_i^{p,h}$ and in formulas (86)-(119) for the annihilation contributions that depend on the coefficients $b_i^{p,h}$. We are only interested here in the decay modes of Table~\ref{tab:B_decays}.\par
\noi To give an example e.g. $\overline{B}_d \to \overline{K}^{*0} \omega$, the first decay in the Table, the amplitude reads
$$A^h(\overline{B}_d \to \overline{K}^{*0} \omega) = {1 \over \sqrt{2}} \{ \left[ \lambda'_u a_2^h + \sum_{p = u, c} \lambda'_p \left (2(a_3^h+a_5^h) + { 1 \over 2} (a_7^{p,h} + a_9^{p,h})\right )\right ] A_{K^*\omega}$$
\beq
\label{12e}
+ \sum_{p = u, c} \lambda'_p \left(a_4^{p,h} - {1 \over 2} a_{10}^{p,h} \right) A_{\omega K^*} \} + {1 \over \sqrt{2}} \left(\lambda'_u+\lambda'_c \right) \left(b_3^h - {1 \over 2} b_3^{EWh} \right) B_{K^* \omega}
\eeq

\noi The expressions for the coefficients $a_i^{p,h}, b_i^h$ are given in Appendices B and C. By inspection of these formulas one can see that the coefficients $a_i^{p,h} (h = 0, -)$ depend on the initial and final states and will therefore depend on the hadronic parameters relevant for each decay, given in the tables~\ref{tab:3} and~\ref{tab:4}. Therefore, strictly speaking, the parameters $a_i^{p,h}$ and $b_i^h$ are not constants, and will be functions of these hadronic parameters, e.g. $a_2^h \to a_2^h(\overline{B}_d, \overline{K}^{*0}, \omega)\ (h = 0, -)$, and so on for the other coefficients and the other decay modes.\par
\noi Therefore, for a given coefficient with a given helicity $a_i^{p,h} (h = 0, -)$ there is a dispersion of its value due to the different decays. According to the formulas (49)-(76) of \cite{BUCHALLA}, these coefficients contribute to a number of decays, for which, using the formulas of Appendices B and C for $a_i^{p,h}, b_i^h$ and the central values for the hadronic parameters of tables~\ref{tab:3} and~\ref{tab:4}, we find their dispersion as can be seen in Tables~\ref{tab:5} and~\ref{tab:6}. 

\begin{table}
\centering
\footnotesize
\begin{tabular}{|c|c|c|c|}
\hline 
Coefficient $a_i^0$ & Re($a_i^0$) & Im($a_i^0$) & $\sigma_g(a_i^0)$ \\ \hline \hline
$a_1^0$ & $0.945 \pm 1\%$ & $0.014 \pm 0\%$ & $\pm 10\%$ \\ \hline
$a_2^0$ & $0.302 \pm 25\%$ & $-0.081 \pm 0\%$ & $\pm 10\%$ \\ \hline
$a_3^0$ & $-0.008 \pm 50\%$ & $0.003 \pm 0\%$ & $\pm 10\%$ \\ \hline
$a_4^{0u}$ & $-0.021 \pm 7\%$ & $-0.014 \pm 0\%$ & $\pm 10\%$ \\ \hline
$a_4^{0c}$ & $-0.029 \pm 5\%$ & $-0.009 \pm 0\%$ & $\pm 10\%$ \\ \hline
$a_5^0$ & $0.015 \pm 33\%$ & $-0.003 \pm 0\%$ & $\pm 10\%$ \\ \hline
$a_7^{0u}/\alpha$ & $-0.271 \pm 110\%$ & $-0.680 \pm 98\%$ & $\pm 10\%$ \\ \hline
$a_7^{0c}/\alpha$ & $0.020 \pm 80\%$ & $0.004 \pm 0\%$ & $\pm 10\%$ \\ \hline
$a_9^{0u}/\alpha$ & $-1.365 \pm 22\%$ & $-0.680 \pm 95\%$ & $\pm 10\%$ \\ \hline
$a_9^{0c}/\alpha$ & $-1.058 \pm 2\%$ & $-0.018 \pm 0\%$ & $\pm 10\%$ \\ \hline
$a_{10}^{0u}/\alpha$ & $-0.334 \pm 25\%$ & $0.080 \pm 0\%$ & $\pm 10\%$ \\ \hline
$a_{10}^{0c}/\alpha$ & $-0.340 \pm 23\%$ & $0.083 \pm 0\%$ & $\pm 10\%$ \\ \hline

\end{tabular}

\caption{\label{tab:5}The $a_i^0$ QCDF coefficients for the helicity $h = 0$. $\sigma_g(a_i^0)$ is an additional {\it ad hoc} error to be added in quadrature with the specific errors of $a_i^0$.}

\end{table}

\begin{table}
\centering
\footnotesize
\begin{tabular}{|c|c|c|c|}
\hline 
Coefficient $a_i^-$ & Re($a_i^-$) & Im($a_i^-$) & $\sigma_g(a_i^-)$ \\ \hline \hline
$a_1^-$ & $1.126 \pm 1\%$ & $0.029 \pm 0\%$ & $\pm 10\%$ \\ \hline
$a_2^-$ & $-0.207 \pm 9\%$ & $-0.162 \pm 0\%$ & $\pm 10\%$ \\ \hline
$a_3^-$ & $0.021 \pm 16\%$ & $0.005 \pm 0\%$ & $\pm 10\%$ \\ \hline
$a_4^{-u}$ & $-0.045 \pm 1\%$ & $-0.015 \pm 0\%$ & $\pm 10\%$ \\ \hline
$a_4^{-c}$ & $-0.043 \pm 1\%$ & $-0.001 \pm 0\%$ & $\pm 10\%$ \\ \hline
$a_5^-$ & $-0.026 \pm 15\%$ & $-0.006 \pm 0\%$ & $\pm 10\%$ \\ \hline
$a_7^{-u}/\alpha$ & $1.052 \pm 25\%$ & $0.009 \pm 0\%$ & $\pm 10\%$ \\ \hline
$a_7^{-c}/\alpha$ & $1.024 \pm 25\%$ & $0.009 \pm 0\%$ & $\pm 10\%$ \\ \hline
$a_9^{-u}/\alpha$ & $-0.279 \pm 99\%$ & $-0.037 \pm 0\%$ & $\pm 10\%$ \\ \hline
$a_9^{-c}/\alpha$ & $-0.307 \pm 90\%$ & $-0.037 \pm 0\%$ & $\pm 10\%$ \\ \hline
$a_{10}^{-u}/\alpha$ & $0.292 \pm 21\%$ & $0.171 \pm 0\%$ & $\pm 10\%$ \\ \hline
$a_{10}^{-c}/\alpha$ & $0.293 \pm 21\%$ & $0.182 \pm 0\%$ & $\pm 10\%$ \\ \hline

\end{tabular}

\caption{\label{tab:6}The $a_i^-$ QCDF coefficients for the helicity $h = -$. $\sigma_g(a_i^-)$ is an additional {\it ad hoc} error to be added in quadrature with the specific errors of $a_i^-$.}

\end{table}

\noi For some coefficients the dispersion is small, like in the cases of the important tree coefficient $a_1^h$ or the penguin coefficients $a_4^{p,h}$. In other cases the dispersion is very large, and can reach up to $100 \%$, mostly for the electroweak coefficients, specially for the ones that are affected by long distance corrections, as explained in the appendices B, C.\par 
\noi Since we have used the central hadronic parameters of tables~\ref{tab:3} and~\ref{tab:4}, that are also affected with some errors, we have introduced a general overall error $\sigma_g$.\par

\begin{table}
\centering
\footnotesize
\begin{tabular}{|c|c|c|c|c|}
\hline 
Index $i$ & Re($b_i^0$) & Im($b_i^0$) & Re($b_i^-$) & Im($b_i^-$) \\ \hline \hline
$1$ & $9.692 \pm 1.110$ & $-4.052 \pm 1.060$ & $0.691 \pm 0.040$ & $0 \pm\ 0$ \\ \hline
$2$ & $-3.038 \pm 0.347$ & $1.268 \pm 0.331$ & $-0.0217 \pm 0.013$ & $0 \pm\ 0$ \\ \hline
$3$ & $3.372 \pm 0.915$ & $-3.784 \pm 1.210$ & $-3.736 \pm 0.929$ & $3.855 \pm 1.210$ \\ \hline
$4$ & $-1.203 \pm 0.138$ & $0.503 \pm 0.131$ & $-0.086 \pm 0.005$ & $0 \pm\ 0$ \\ \hline
$3_{EW}$ & $-0.123 \pm 0.012$ & $0.080\pm 0.023$ & $0.0305 \pm 0.0094$ & $-0.0399 \pm 0.0126$ \\ \hline
$4_{EW}$ & $0.035 \pm 0.004$ & $-0.015\pm 0.004$ & $0.0025 \pm 0.0001$ & $0 \pm\ 0$ \\ \hline

\end{tabular}

\caption{\label{tab:7}The $b_i^0$, $b_i^-$ QCDF annihilation coefficients for the helicities $h = 0, -$. }

\end{table}

\noi The QCDF annihilation coefficients $ b_i^h$ with their errors are given in Table~\ref{tab:7}.

\section{Comparison of QCD Factorization with data for $B$ decays into two light vector mesons}
\begin{figure}[hbt]
\vfill
\begin{center}

%\vskip 5cm
%\includegraphics[width=0.8\textwidth]{figures/Figure_1.png}
\includegraphics[width=0.8\textwidth]{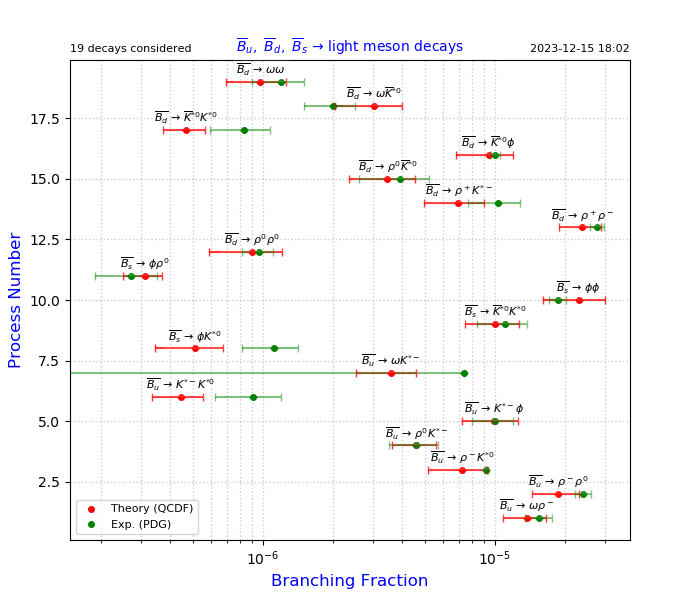}

\caption{\label{fig:Br} The CP-averaged Branching Fractions for the branching ratios for $B$ decays into two light vectors mesons.}
\end{center}
\vfill

\end{figure}

\begin{figure}[hbt]
\vfill
\begin{center}

%\vskip 5cm
%\includegraphics[width=0.8\textwidth]{figures/Figure_2.png}
\includegraphics[width=0.8\textwidth]{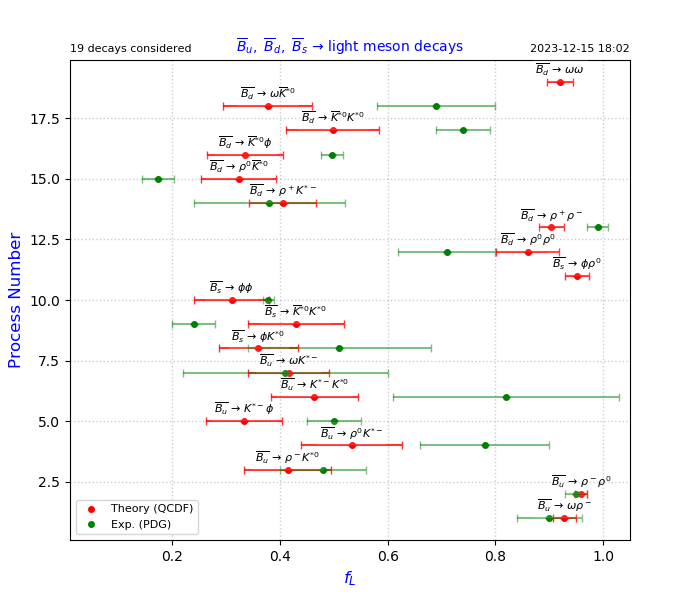}

\caption{\label{fig:fL} The longitudinal fractions, $f_L$, for $B$ decays into two light vectors mesons.}
\end{center}
\vfill

\end{figure}

\begin{figure}[hbt]
\vfill
\begin{center}

%\vskip 5cm
%\includegraphics[width=0.8\textwidth]{figures/Figure_3.png}
\includegraphics[width=0.8\textwidth]{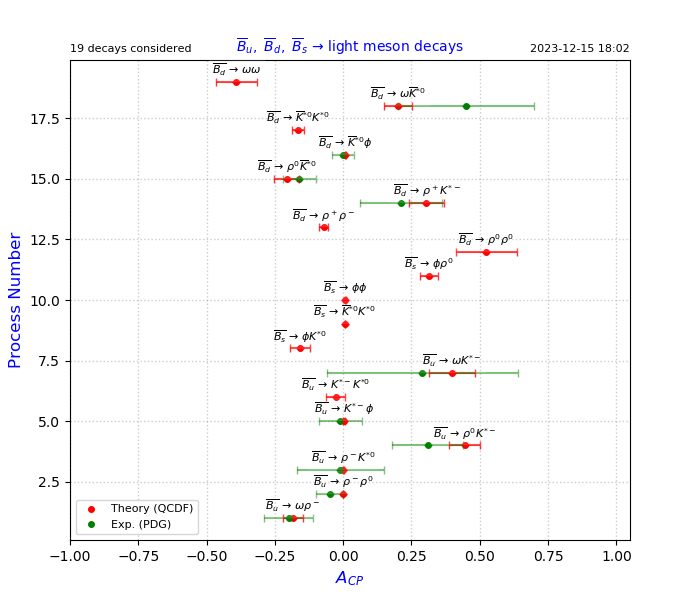}

\caption{\label{fig:Acp} The direct CP violation, $A_{CP}$, for $B$ decays into two light vectors mesons.}
\end{center}
\vfill

\end{figure}

We summarize in the Table~\ref{tab:8} our calculations for the branching fractions, the longitudinal polarizations, $f_L$, and the CP asymmetry, $A_{CP}$, using QCD Factorization for the modes with available data. The values obtained with QCD Factorization are compared to the data in the Figures~\ref{fig:Br} to~\ref{fig:Acp}. In order to quantify the level of agreement between data and QCD Factorization predictions, we have calculated the $\chi^2$.

\subsection{CP-averaged branching ratios}

%%%%%%%%%%%%%%%%%%%% Figure Feynman diagram Bs to DsK  %%%%%%%%%%%%%%%%%%%
%%%%%%%%%%%%%%%%%%%%%%%%% End Figure %%%%%%%%%%%%%%%%%%%%%%%%%

\begin{table}
\centering
\footnotesize

\begin{tabular}{|c|c|c|c|}
\hline 
Decay mode &  $\rm{BR}^{th}\ (\times 10^{-6})$ & $f_L^{th}$ & $A_{CP}^{th}$\\ \hline \hline 
$\overline{B}_d \to \omega \omega$ &  $0.98 \pm 0.28$ & $0.92 \pm 0.03$ & $-0.391 \pm 0.076$  \\ \hline
$\overline{B}_d \to \overline{K}^{*0} \omega$ &  $2.93 \pm 0.95$ & $0.387 \pm 0.085$ & $0.187 \pm 0.047$  \\ \hline
$\overline{B}_d \to \overline{K}^{*0} K^{*0}$ &  $0.47 \pm 0.10$ & $0.498 \pm 0.086$ & $-0.165 \pm 0.022$ \\ \hline
$\overline{B}_d \to \overline{K}^{*0} \phi$ & $9.29 \pm 2.55$ & $0.340 \pm 0.071$ & $0.009 \pm 0.002$ \\ \hline
$\overline{B}_d \to \overline{K}^{*0} \rho^0$ &  $3.45 \pm 1.11$ & $0.324 \pm 0.070$ & $-0.189 \pm 0.040$ \\ \hline
$\overline{B}_d \to \overline{K}^{*-} \rho^+$ & $6.97 \pm 2.00$ & $0.405 \pm 0.062$ & $0.305 \pm 0.065$ \\ \hline
$\overline{B}_d \to \rho^- \rho^+$ &  $23.84 \pm 4.91$ & $0.904 \pm 0.023$ & $-0.071 \pm 0.015$ \\ \hline
$\overline{B}_d \to \rho^0 \rho^0$ &  $0.73 \pm 0.20$ & $0.841 \pm 0.061$ & $0.572 \pm 0.099$ \\ \hline \hline
$\overline{B}_s \to \phi \rho^0$ & $0.31 \pm 0.06$ & $0.951 \pm 0.022$ & $0.314 \pm 0.033$ \\ \hline
$\overline{B}_s \to \phi \phi$ & $22.63 \pm 6.67$ & $0.315 \pm 0.070$ & $0.007 \pm 0.002$ \\ \hline$\overline{B}_s \to \overline{K}^{*0} K^{*0}$ &  $10.05 \pm 2.64$ & $0.429 \pm 0.088$ & $0.006 \pm 0.001$ \\ \hline
$\overline{B}_s \to K^{*0} \phi$ & $0.50 \pm 0.16$ & $0.365 \pm 0.074$ & $-0.160 \pm 0.037$ \\ \hline \hline
$B^- \to K^{*-} \omega$ &  $3.43 \pm 0.99$ & $0.419 \pm 0.076$ & $0.408 \pm 0.087$ \\ \hline
$B^- \to K^{*-} K^{*0}$ & $0.44 \pm 0.11$ & $0.464 \pm 0.081$ & $-0.027 \pm 0.034$ \\ \hline
$B^- \to K^{*-} \phi$ &  $9.80 \pm 2.63$ & $0.337 \pm 0.071$ & $0.003 \pm 0.034$ \\ \hline
$B^- \to K^{*-} \rho^0$  & $4.56 \pm 1.00$ & $0.535 \pm 0.093$ & $0.435 \pm 0.0054$ \\ \hline
$B^- \to \overline{K}^{*0} \rho^-$ & $7.24 \pm 2.07$  & $0.413 \pm 0.082$ & $-0.000 \pm 0.001$ \\ \hline
$B^- \to \rho^0 \rho^-$ & $17.92 \pm 3.91$ & $0.956 \pm 0.012$ & $-0.000 \pm 0.001$ \\ \hline
$B^- \to \omega \rho^-$ & $13.14 \pm 2.72$ & $0.924 \pm 0.022$ & $-0.189 \pm 0.037$ \\ \hline
\end{tabular}

\caption{\label{tab:8}The QCDF predictions for the branching ratios, $f_L$ and $A_{CP}$ for $B$ decays into two light vectors mesons.}

\end{table}

\noi We find a good agreement with the data, namely a ${\rm reduced}\ \chi^2 = 0.74$ for 18 measurements, giving a probability of $\sim 77 \%$.

\subsection{Longitudinal polarization fractions}

%%%%%%%%%%%%%%%%%%%% Figure Feynman diagram Bs to DsK  %%%%%%%%%%%%%%%%%%%
%%%%%%%%%%%%%%%%%%%%%%%%% End Figure %%%%%%%%%%%%%%%%%%%%%%%%%
We find an agreement with the data that is not good, namely a ${\rm reduced}\ \chi^2 = 2.62$ for 17 measurements of $f_L$, giving a probability of $\sim 0.03 \%$. However the main disagreement is due to $\rm B_{d,s}\to K^{*0}\overline{K}^{*0}$ and $B^0\to \rho^+\rho^-$. The issues with the modes $\rm B_{d,s}\to K^{*0}\overline{K}^{*0}$ have been identified and is discussed in detail in~\cite{AO:1}. Should one remove these 3 modes, the $\chi^2$ becomes 1.93 for 14 measurements, which means a probability of 2\% that, although not good, is acceptable.

\subsection{Direct CP violation}
%%%%%%%%%%%%%%%%%%%% Figure Feynman diagram Bs to DsK  %%%%%%%%%%%%%%%%%%%
%%%%%%%%%%%%%%%%%%%%%%%%% End Figure %%%%%%%%%%%%%%%%%%%%%%%%%
The direct CP violation, $A_{CP}$,  is defined as:
\beq
\label{13e}
A_{CP} = {\sum_{h=0,-}\left[\Gamma (\overline{B} \to \overline{f}, h) - \Gamma (B \to f, h)\right] \over \sum_{h=0,-}\left[\Gamma (\overline{B} \to \overline{f}, h) + \Gamma(B \to f, h)\right]}
\eeq

\vskip 5 truemm

\noi We find a good agreement with the data, namely a ${\rm reduced}\ \chi^2 = 0.54$ for 10 measurements of $A_{CP}$, giving a probability of $\sim 86 \%$.

\section{Conclusion}

Although not totally satisfactory, QCD Factorization is able to reproduce the available data for B decays to 2 light vector mesons in a reasonably well but also raises some noticeable issues. On the positive side, we find an general agreement with the data for the branching ratios that is good, namely a reduced $\chi^2 = 0.74$ for 18 measurements, giving a probability of $\sim 77 \%$. Similarly, the situation is also good concerning the direct CP violation $A_{CP}$, as we find a reduced $\chi^2 = 0.54$ for 10 measuremeents, giving a probability of $\sim 86 \%$. On the negative side,  the agreement with all the available data is not good for the longitudinal fraction $f_L$. Indeed the reduced $\chi^2$ is $2.62$ for 17 measurements, giving only a probability of $0.03 \%$. The disagreement is mainly due to 3 modes : $\rm B_{d,s}\to K^{*0}\overline{K}^{*0}$ and $B^0\to \rho^+\rho^-$ .  We underline that, in these cases, we have found disagreements between theory and data that are rather pathological. For example there is a very strong U-spin violation in the comparison of the two modes $\overline{B}_{s,d} \to K^{*0} \overline{K}^{*0}$, which we analyze further in detail in a separate paper~\cite{AO:1}. Although our calculations are based on well-established previous works we describe the formalism in some detail in the Appendices in order to facilitate possible future verifications of our numerical results by interested authors.

\newpage
\appendix
\noi {\Large \bf Appendix }
\section{\Large \bf  Wilson operators and short distance coefficients at the different scales}

The effective Lagrangian is given by
\beq
H = {G_F \over \sqrt{2}} \left[V^*_{cs} V_{cb} (C_1 O_1 + C_2 O_2) - V^*_{ts} V_{tb} (C_3 O_3 + C_4 O_4  + C_5 O_5 + C_6 O_6) \right] + H_{EW}^{Penguin}
\label{A-1e}
\eeq
$$O_1 = \left[\overline{c} \gamma^\mu (1-\gamma_5) b \right] \left[\overline{s} \gamma_\mu  (1-\gamma_5) c \right]$$
$$O_2 = \left[\overline{c}_\alpha \gamma^\mu (1-\gamma_5) b_\beta \right] \left[\overline{s}_\beta \gamma_\mu  (1-\gamma_5) c_\alpha \right]$$
$$O_3 = \left[\overline{s} \gamma^\mu (1-\gamma_5) b \right] \left[\overline{q}\gamma_\mu  (1-\gamma_5) q \right]$$
\beq
O_4 = \left[\overline{s}_\alpha \gamma^\mu (1-\gamma_5) b_\beta \right] \left[\overline{q}_\beta \gamma_\mu  (1-\gamma_5) q_\alpha \right]
\label{A-2e}
\eeq
$$O_5 = \left[\overline{s} \gamma^\mu (1-\gamma_5) b \right] \left[\overline{q}\gamma_\mu  (1+\gamma_5) q \right]$$
$$O_6 = \left[\overline{s}_\alpha \gamma^\mu (1-\gamma_5) b_\beta \right] \left[\overline{q}_\beta \gamma_\mu  (1+\gamma_5) q_\alpha \right]$$

\noi where the Penguin contribution reads
\beq
H_{EW}^{Penguin} = - {G_F \over \sqrt{2}} V^*_{ts} V_{tb} (C_7 O_7 + C_8 O_8  + C_9 O_9 + C_{10} O_{10})
\label{A-3e}
\eeq
$$O_7 = {3 \over 2} \left[\overline{s} \gamma^\mu (1-\gamma_5) b \right] \left[e_q \overline{q}\gamma_\mu  (1+\gamma_5) q \right]$$
\beq
O_8 = {3 \over 2} \left[\overline{s}_\alpha \gamma^\mu (1-\gamma_5) b_\beta \right] \left[e_q \overline{q}_\beta \gamma_\mu  (1+\gamma_5) q_\alpha \right]
\label{A-4e}
\eeq
$$O_9 = {3 \over 2} \left[\overline{s} \gamma^\mu (1-\gamma_5) b \right] \left[e_q \overline{q}\gamma_\mu  (1-\gamma_5) q \right]$$
$$O_{10} = {3 \over 2} \left[\overline{s}_\alpha \gamma^\mu (1-\gamma_5) b_\beta \right] \left[e_q \overline{q}_\beta \gamma_\mu  (1-\gamma_5) q_\alpha \right]$$

\noi and the Wilson coefficients are given in Table~\ref{tab:11}.

\begin{table}[b]
\centering
\small

\begin{tabular}{|c|c|c|c|c|}
\hline 
$C_1$ & $C_2$ & $C_3$& $C_4$ & $C_5$\\ \hline
$1.086$ & $-0.191$ & $0.014$ & $-0.035$ & $0.010$ \\ \hline \hline
$C_6$ & $C_7$ & $C_8$ & $C_9$ & $C_{10}$ \\ \hline
$-0.043$ & $0.011\ \alpha$ & $0.059\ \alpha$ & $-1.229\ \alpha$ & $0.246\ \alpha$ \\ \hline
\end{tabular}

\caption{\label{tab:11}  Wilson coefficients at NLO for $\mu = m_b = 4.2\ {\rm GeV}$ in the NDR scheme.}
\end{table}

\subsection {\bf \large Different scales in QCD Factorization}

Let us make an important remark about the scales considered in QCDF.\par 
\noi The Wilson coefficients $C_i(\mu)$ relevant to the Vertex $V$ and Penguin $P$ contributions and the order $\alpha_s(\mu)$ NLO corrections specific to QCDF are estimated at some scale $\mu$, that for definiteness we take to be $\mu = m_b = 4.2\ {\rm GeV}$.\par
\noi For the Hard scattering $H$, due to off-shellness of the gluon in these diagrams \cite{BBNS-01}, one adopts {\it a different scale} $\mu_h = \sqrt{\Lambda_h \mu}$ with $\Lambda_h = 0.5\ {\rm GeV}$ and $\mu = m_b$, evaluating both at this scale the Wilson coefficients $C_i(\mu_h)$ and the coupling $\alpha_s(\mu_h)$ specific to the NLO corrections of QCDF. For the Annihilation corrections, one adopts the same Wilson coefficients $C_i(\mu_h)$ and coupling $\alpha_s(\mu_h)$.\par

\noi This important point, that has sizeable quantitative consequences, has been put forward in refs. \cite{BENEKE-2007} and \cite{BUCHALLA}, following the earlier paper by Beneke and Neubert \cite{BENEKE-NEUBERT}, who underline that both the running coupling constant and the Wilson coefficients for the Hard scattering and Annihilation terms have to be evaluated at an intermediate scale $\mu_h \sim \sqrt{\Lambda_{QCD} m_b}$ rather than $\mu \sim m_b$.\par

\noi This different scale $\mu_h$ is much lower than $\mu$, namely
\beq
\label{C-7-2e}
\mu = m_b = 4.2\ {\rm GeV} \ , \qquad \Lambda_h = 0.5\ {\rm GeV} \ , \qquad \mu_h = \sqrt{\mu \Lambda_h} = 1.45\ {\rm GeV} 
\eeq

\noi This increases substantially $\alpha_s(\mu_h)$ and also the Wilson coefficients evaluated at this scale. One has 
\beq
\label{C-7-3e}
\alpha_s(4.2) = 0.224 \ , \qquad \qquad \alpha_s(1.45) = 0.338
\eeq

\noi and the Wilson coefficients at this latter scale are given in Table~\ref{tab:12} (for the formalism see the detailed paper \cite{BUCHALLA-HARLANDER-1995}). \par

\begin{table}[h]
\centering
\small

\begin{tabular}{|c|c|c|c|c|}
\hline 
$C_1$ & $C_2$ & $C_3$& $C_4$ & $C_5$\\ \hline
$1.190$ & $-0.373$ & $0.027$ & $-0.062$ & $0.012$ \\ \hline \hline
$C_6$ & $C_7$ & $C_8$ & $C_9$ & $C_{10}$ \\ \hline
$-0.086$ & $0.003\ \alpha$ & $0.128\ \alpha$ & $-1.353\ \alpha$ & $0.430\ \alpha$ \\ \hline
\end{tabular}
\caption{\label{tab:12} Wilson coefficients at NLO in the NDR scheme, at the scale $\mu_h = \sqrt{\mu \Lambda_h} = 1.45\ {\rm GeV}$, relevant for the Hard scattering and Annihilation contributions.}
\end{table}

\noi Clearly, the Hard scattering and Annihilation diagrams will be very much enhanced due to the low scale $\mu_h$.

\section{\Large \bf  The coefficients $a_i^h$}

\noi The normalization factors satisfy
\beq
\label{B-2e}
N_i^h(\phi) = 0\ \ (i = 6, 8) \ , \qquad N_i^h(\phi) = 1\ \ (i \not = 6, 8)\ , \qquad (h = 0, -)
\eeq

\subsection {\large \bf  Longitudinal $h = 0$ polarization coefficients $a_i^0$}

From formulas (11) of \cite{KAGAN}, (53) of \cite{BENEKE-2007} and (24) of \cite{BUCHALLA} the relevant coefficients $a_i^{0,p} (p = u, c)$ of QCD Factorization write, for $i = 1, 2, 3, 4, 5, 7, 9, 10$,

$$a_i^{0,p} = \left(C_i(\mu) + {C_{i\pm 1}(\mu) \over N_c} \right) N_i^0 + {C_{i\pm 1}(\mu) \over N_c} {C_F \alpha_s(\mu) \over {4 \pi}}\ V_i^0$$
$$- \ \left({C_{i+1}(\mu) \over N_c} \delta_{i,4} + {C_{i-3}(\mu) \over N_c} \delta_{i,10}\right){C_F \alpha_s(\mu) \over {4 \pi}} r_\perp(\mu)\ V_ \perp^0$$
\beq
\label{B-1e} + {C_{i\pm 1}(\mu_h) \over N_c} {C_F \alpha_s(\mu_h) \over {4 \pi}} {4 \pi^2 \over N_c}\ H_i^0 + P_i^{0,p}(\mu)
\eeq

\noi where the upper (lower) signs apply when $i$ is odd (even) and we have made explicit that the scale is not the same in front of the Hard scattering term $H_i^0$. The coupling $\alpha_s$ and the Wilson coefficients for both scales are given respectively by (\ref{C-7-3e})  and Tables~\ref{tab:11} and~\ref{tab:12}.\par

\subsubsection {\bf Vertex corrections}

For $\mu = m_b$ one gets the simplified formulas \cite{BUCHALLA},

$$\ \ V_i^0 = - 18 + \int_0^1 dy\ \phi_\parallel (y) g(y) \ , \qquad \qquad i \in \{1, 2, 3, 4, 9, 10 \}$$
\beq
\label{B-3e}
\ \ \ V_i^0 = 6 - \int_0^1 dy\ \phi_\parallel(y) g(1-y) \ , \qquad \ \ \ \ i \in \{5, 7 \} \qquad \qquad
\eeq

\vskip 3 truemm

\noi with
$$g(y) = 3 \left( {1-2y \over 1-y} \ln y - i \pi \right)$$ 
\beq
\label{B-4e}
+ \left (2 {\rm Li}_2(y) - \ln^2 y + {2 \ln y \over 1-y} - (3 + 2 \pi i) \ln y - [y \to 1-y ] \right)
\eeq

\vskip 3 truemm

\noi and the lower twist distribution functions read
\beq
\label{B-5bise}
\phi_\parallel (x) = 6x(1-x) \ , \qquad \qquad \phi_\perp (x) = 6x(1-x)
\eeq
\beq
\label{B-6bise}
\phi_v(x) = \int^x_0 du {\phi_\perp(u) \over 1-u} - \int^1_x du {\phi_\perp(u) \over u} = 3(2x-1)
\eeq

\vskip 3 truemm

\noi Numerically one finds

$$V_i^0 = - 18 - {1 \over 2} - 3\pi i , \ \qquad \qquad i \in \{1, 2, 3, 4, 9, 10 \}$$
\beq
\label{B-6-bise}
V_i^0 = 6 + {1 \over 2} - 3 \pi i  \ , \qquad \qquad \ \ \ \ i \in \{5, 7 \} \qquad \qquad
\eeq

\vskip 3 truemm

\noi Finally, the perpendicular vertex correction $V_\perp^0$ is given by \cite{BUCHALLA}
\beq
\label{B-7bise} 
V_\perp^0 = \int_0^1 dx\ h(x)\ \phi_v(x)
\eeq

\noi where the function $h(x)$ is given by
\beq
\label{B-4bise}
h(x) = 2 {\rm Li}_2(x) - \ln^2 x - (1 + 2 \pi i) \ln x - [x \to 1-x ]
\eeq

\noi and numerically one finds
\beq
\label{B-4tere}
V_\perp^0 = 9 - 6 \pi i
\eeq

\subsubsection {\bf  Penguin corrections} 

The non-vanishing $h = 0$ Penguin corrections $P_i^{p,0}(\mu) (p = u, c)$ in the general formula (\ref{B-1e}) are given in terms of functions computed in ref. \cite{BUCHALLA}, that we reproduce here for the sake of completeness. One has
\beq
\label{PE-2e}
P_4^{0,p} = {C_F \alpha_s \over {4 \pi N_c}} \left(P_2^p - r_\perp P_3^p \right)
\eeq
\beq
\label{PE-3e}
P_7^{0,p} = P_9^{p,0} = {\alpha \over {9\pi}} P_n^{p,EW}
\eeq
\beq
\label{PE-4e}
P_{10}^{0,p} =  {\alpha \over {9\pi N_c}} \left(P_2^{p,EW}- r_\perp P_3^{p,EW} \right)
\eeq

\noi The functions $P_2^p, P_2^{p,EW}, P_3^p, P_3^{p,EW}, P_n^{p,EW}$ are given as follows.\par 
\noi The penguin contributions from the twist 2 function (\ref{B-5bise}) read for $\mu = m_b$
$$P^p_2 = C_1\left[{2 \over 3} - G_V(s_p)\right] + C_3 \left[{4 \over 3} - G_V(0) - G_V(1)\right]$$ 
$$+\ (C_4 +C_6) \left[-3G_V(0)-G_V(s_c)-G_V(1)\right] - 6 C_{8g}$$
\beq
\label{Pe}
P^{p,EW}_2 = (C_1 + N_c C_2) \left[{2 \over 3} - G_V(s_p)\right] - 9 C_{7\gamma}
\eeq

\noi The function $G_V(s)$ reads
\beq
\label{B-9e}
G_V(s) = \int^1_0 dy\ \phi_\parallel(y) G(s-i\epsilon,1-y)
\eeq

\noi where $G(s,y)$ is the penguin function, formula (51) of \cite{BBNS-01},
\beq
\label{B-10bise}
G(s,x) = {2(12s + 5x - 3x\ln s) \over 9x} - {4 \sqrt{4s-x} (2s+x) \over 3 x^{3/2}} \arctan\sqrt{{x \over 4s-x}}
\eeq

\noi From $m_c = 1.3\ {\rm GeV}, m_b = 4.2\ {\rm GeV}$ one has $s_c = 0.958$ and the values 
\beq
\label{B-11bise}
G_V(0) = {5 \over 3} + {2\pi i \over 3} \ , \qquad G_V(1) = {85 \over 3} - 6\sqrt{3}\pi + {4\pi^2 \over 9}\ , \qquad G_V(s_c) = 2.316+1.044\ i 
\eeq

\noi The penguin contributions from the twist-3 terms are given by
$$P_3^p = -\left[ C_1 {\hat G}(s_p) + C_3\left({\hat G}(0)+{\hat G}(1)\right) + (C_4+C_6)\left( 3{\hat G}(0)+{\hat G}(s_c)+{\hat G}(1) \right) \right]$$
\beq
\label{P-1e}
P_3^{p,EW} = - \left( C_1 + N_c C_2\right) {\hat G}(s_p)
\eeq

\noi where the function ${\hat G}(s)$ is given by 
\beq
\label{P-2e}
{\hat G}(s) = \int^1_0 dy\ \phi_v(y) G(s-i\epsilon,1-y)
\eeq
\noi with $\phi_v$ given by~\ref{B-6bise}.
\noi From (\ref{P-2e}) and (\ref{B-10bise}) one gets 
$$ {\hat G}(s) = 1 - 36 s_c + 12 s_c \sqrt{1-4s_c} \left( 2 {\rm arctanh} \sqrt{1-4s_c} - i\pi \right) - 12 s_c^2 \left( 2 {\rm arctanh} \sqrt{1-4s_c} - i\pi \right)^2 \ ,$$
\beq
\label{P-3e}
{\hat G}(0) = 1 \ , \qquad \qquad \qquad {\hat G}(1) = {4 \over 3}\ \pi^2 + 4 \sqrt{3} \pi - 35
\eeq

\noi The electromagnetic penguin contributions, with the gluon replaced by a photon, are given by for $\mu = m_b$,
\beq
\label{P-4e}
P_n^{c,EW} = \left(C_1 + N_c C_2 \right) \left({2 \over 3} + {4 \over 3} \ln {m_c \over m_b} \right) - 3C_{7\gamma}^{eff}
\eeq

\noi In the case of the $u$-quark one needs a long-distance model for the light-quark loop, and one proposes for  $\mu = m_b$ \cite{BUCHALLA},
$$P_n^{u,EW}(V) = \left(C_1 + N_c C_2 \right) (- {10 \over 9} + {4\pi^2 \over 3} \sum_{r=\rho,\omega} {f_r^2 \over {m_V^2 - m_r^2 + i m_r \Gamma_r}} $$
\beq
\label{P-5e}
- {2 \pi \over 3} {m_V^2 \over t_c}\ i + {2 \over 3} \ln {m_V^2 \over m_b^2} + {2 \over 3} {t_c-m_V^2 \over t_c} \ln {t_c-m_V^2 \over m_V^2} ) - 3C_{7\gamma}^{eff}
\eeq

\noi with $t_c = 4 \pi^2 \left(f_\rho^2 + f_\omega^2 \right)$.

\subsubsection {\bf  Hard scattering corrections} 

The non-vanishing Hard Scattering corrections and the Weak annihilation contributions are estimated at the scale $\mu = \sqrt{m_b\Lambda_h} = 1.45$. The longitudinal Hard Scattering contributions $H_i^0$ in (\ref{B-1e}) are given by the expressions
$$H_i^0 = {f_B f_V \over m_B^2 A_0^{B \to V}(0)} {m_B \over \lambda_B}\ 9 \left[1 + r_\perp^V (X_H-2)\right] \ , \ \ \ \ \qquad i \in \{1, 2, 3, 4, 9, 10 \}$$
\beq
\label{B-13e}
H_i^0 = -  {f_B f_V \over m_B^2 A_0^{B \to V}(0)} {m_B \over \lambda_B}\ 9 \left[1 + r_\perp^V (X_H-2)\right]\ , \qquad \qquad i \in \{5, 7 \}
\eeq
\noi where
$$\int_0^1 {d\xi \over \xi} \phi_B(\xi) = {m_B \over \lambda_B}$$

\noi and $r_\perp^V$ is given by
\beq
\label{H-1e}
r_\perp^V = {2m_V f_\perp^V \over m_b f_V}
\eeq

\vskip 3 truemm

\noi Let us make a few remarks on these formulas.\par
\noi The quantities $H_i^0$ are pieces of the coefficients $a_i^0$ (\ref{B-1e}) inversely proportional to the form factor $A_0^{B \to V}(0)$. Notice that this gives a form-factor dependence of the coefficients $a_i^0$, the rest of the terms in (\ref{B-1e}) being form factor independent.\par  
\noi The parameter $r_\perp^V$ (\ref{H-1e}) is scale-dependent, and we need the parameters $r_\perp^V$ at the scale $\mu_h = 1.45\ {\rm GeV}$,  
\beq
\label{H-2}
r^V_\perp(\mu_h) = {2m_V f_V^\perp(2{\rm GeV}) \over m_b f_V} \left[{\alpha_s(\mu_h) \over \alpha_s(m_b)} \right]^{-3C_F/\beta_0} \left[{\alpha_s(\mu_h) \over \alpha_s (2\ GeV)} \right]^{C_F/\beta_0}
\eeq
\noi For example, in the $\rho$ case one has $f_\rho^\perp(2{\rm GeV}) = 0.150\ {\rm GeV}$ and one finds $r^\rho_\perp(\mu_h) = 0.219$.
\noi Finally, we take for the parameter $\lambda_B = 0.2\ {\rm GeV}$ and $X_H$ parametrizes a divergent integral, as given by (64) of \cite{BENEKE-2007}, for which we take
\beq
\label{C-9-1bise}
X_H = {\rm ln}\left( {m_B \over \Lambda_h}\right)
\eeq

\noi with $\Lambda_h = 0.5\ {\rm GeV}$.
\vskip 3 truemm

\noi {\large \bf Summary on the coefficients $a_i^0$}

\noi Gathering the precedent results in (\ref{B-1e}), and taking care of the two different scales $\mu = m_b$ and $\mu_h = \sqrt{\mu \Lambda_h}$ with $\Lambda_h = 0.5\ {\rm GeV}$ for the different terms, one finds the central values of the coefficients $a_i^0$ of Table~\ref{tab:13}.

{\small
\begin{table}[h]
\centering
\small

\begin{tabular}{|c|c|c|}
\hline 
Coefficient & ${\rm Re}(a_i^0)$ & ${\rm Im}(a_i^0)$\\ \hline \hline 
$a_1^0$ & $0.945$ & $0.014$ \\ \hline
$a_2^0$ & $0.302$ & $-0.081$ \\ \hline
$a_3^0$ & $-0.008$ & $0.003$ \\ \hline
$a_4^{0,u}$ & $-0.021$ & $-0.014$ \\ \hline
$a_4^{0,c}$ & $-0.029$ & $-0.009$ \\ \hline
$a_5^0$ & $0.015$ & $-0.003$ \\ \hline
$a_7^{0,u}/\alpha$ & $-0.271$ & $-0.680$ \\ \hline
$a_7^{0,c}/\alpha$ & $0.020$ & $0.004$ \\ \hline
$a_9^{0,u}/\alpha$ & $-1.365$ & $-0.680$ \\ \hline
$a_9^{0,c}/\alpha$ & $-1.058$ & $-0.018$ \\ \hline
$a_{10}^{0,u}/\alpha$ & $-0.334$ & $0.080$ \\ \hline
$a_{10}^{0,c}/\alpha$ & $-0.340$ & $0.084$ \\ \hline
\end{tabular}

\caption{\label{tab:13}  Central values of the coefficients $a_i^0$.}
\end{table} }

\noi The values in the table are central values but one must take into account that these coefficients vary from one mode to another. We take into account the dispersion of these coefficients due to the different modes giving a central value and an error in Table~\ref{tab:5}.

\subsection {\large \bf Transverse $h = -$ polarization coefficients $a_i^-$}

From formulas (11) of \cite{KAGAN}, (53) and (16) of \cite{BENEKE-2007} the coefficients $a_i^{-,p} (p = u, c)$ of QCD Factorization write, for $i = 1, 2, 3, 4, 5, 7, 9, 10$,

$$a_i^{-,p} = \left(C_i(\mu) + {C_{i\pm 1}(\mu) \over N_c} \right) N_i^- + {C_{i\pm 1}(\mu) \over N_c} {C_F \alpha_s(\mu) \over {4 \pi}}\ V_i^-$$
$$- \ \left({C_{i+1}(\mu_h) \over N_c} H_{i+2}^- \delta_{i,4} + {C_{i-3}(\mu_h) \over N_c} H_{i-2}^-  \delta_{i,10}\right){C_F \alpha_s(\mu_h) \over {4 \pi}} {4 \pi^2 \over N_c} r_\perp(\mu_h)$$
\beq
\label{BB-1e} + {C_{i\pm 1}(\mu_h) \over N_c} {C_F \alpha_s(\mu_h) \over {4 \pi}} {4 \pi^2 \over N_c}\ H_i^- + P_i^{-,p}(\mu)
\eeq

\vskip 4 truemm

\noi where the upper (lower) signs apply when $i$ is odd (even) and we have made explicit that the scale is not the same in front of the Hard scattering term $H_i^-$. The coupling $\alpha_s$ and the Wilson coefficients for both scales are given respectively by (\ref{C-7-3e})  and Tables~\ref{tab:11} and~\ref{tab:12}.\par

\subsubsection {\bf  Vertex corrections}

 For the non-vanishing contributions, for $\mu = m_b$ one gets, the simplified formulas

$$\ \ \ \ \ V_i^- = \int_0^1 dy \phi_{b}(y) \left[- 18 + g_T(y) \right] \ , \qquad \qquad i \in \{1, 2, 3, 4, 9, 10 \}$$
\beq
\label{B-3e}
\qquad V_i^- = \int_0^1 dy \phi_{a}(y) \left[6 - g_T(y) \right] \ , \qquad \qquad \ \ \ \ i \in \{5, 7 \} \qquad \qquad
\eeq

\vskip 3 truemm

\noi with
$$g_T(y) = {4-6 y \over 1-y} \ln y - 3 i \pi$$ 
\beq
\label{B-4e}
+ \left (2 {\rm Li}_2(y) - \ln^2 y + {2 \ln y \over 1-y} - (3 + 2 \pi i) \ln y - [y \to 1-y ] \right)
\eeq

\vskip 3 truemm

\noi Keeping only the lower twist, the distribution functions read
\beq
\label{B-5e}
\Phi_V(v) = 6v(1-v)
\eeq
\beq
\label{B-6e}
\phi_a(u) = \int^1_u dv {\Phi_V(v) \over v}  = 3(1-u)^2\ , \qquad \phi_b(u) = \int^u_0 dv {\Phi_V(v) \over 1-v} = 3u^2
\eeq

\vskip 3 truemm

\noi Numerically, one gets, for $\mu = m_b$,
$$\ V_i^-(\phi) = -14 - 6i\pi \ , \qquad \qquad i \in \{1, 2, 3, 4, 9, 10 \}$$
\beq
\label{B-7e}
V_i^-(\phi) = 2 + 6i\pi \ , \qquad \qquad \ \ \ \ \ i \in \{5, 7 \} \qquad \qquad
\eeq

\subsubsection {\bf Penguin corrections} 

Formulas (57)-(59) of \cite{BENEKE-2007} simplify for $\mu = \nu = m_b$, 
$$P_4^{-,p} = {\alpha_s C_F \over 4 \pi N_c}\ \{ C_1 \left[{2 \over 3} - G^-(s_p) \right] + C_3 \left[{4 \over 3} - G^-(0) - G^-(1)\right]$$
\beq
\label{B-8e}
+\ (C_4 +C_6) \left[-3 G^-(0) - G^-(s_p) - G^-(1)\right] \}
\eeq
$$P_7^{-,p} = P_9^{-,p} = - {\alpha \over 3 \pi}\ C_{7\gamma}^{eff} {m_B^2 \over m_{V_2}^2} + {2 \alpha \over 27 \pi} (C_1 + N_c C_2)\left[\delta_{pc} \ln {m_c^2 \over m_b^2} +1 \right]$$
$$P_{10}^{-,p} = {\alpha \over 9 \pi N_c} (C_1 + N_c C_2) \left[{2 \over 3} - G^-(s_p) \right]$$

\noi The enhancement factor ${m_B^2 \over m_{V_2}^2}$ modifies the naive power counting in the EW penguins.\par
\noi The function $G^-(s)$ reads
\beq
\label{B-9e}
G^-(s) = \int^1_0 dy\ \phi_b(y) G(s-i\epsilon,1-y)
\eeq

\noi where $G(s,y)$ is the penguin function (\ref{B-10bise}).

\noi From $m_c = 1.3\ {\rm GeV}, m_b = 4.2\ {\rm GeV}$ one has $s_c = 0.958$ and the values 
\beq
\label{B-11e}
G^-(0) = 2.333+2.094\ i \ , \qquad G^-(1) = 0.035 \ , \qquad G^-(s_c) = 2.090+0.315\ i 
\eeq

\noi From (\ref{B-11e}), and the value \cite{BBNS-01}
\beq
\label{B-12e}
C_{7\gamma}^{eff}(m_b) = -0.318
\eeq

\noi we find for the Penguin contributions at $\mu = m_b$,  
$$P_4^{-,u} = -0.0086 - 0.0131 i \ , \qquad \qquad P_4^{-,c} = -0.0067 + 0.0012 i$$
\beq
\label{B-12e}
P_7^{-,u} = P_9^{-,u} = 0.0073 \ , \qquad \qquad  P_7^{-,c} = P_9^{-,c} = 0.0071
\eeq
$$P_{10}^{-,u} = P_{10}^{-,c} = (- 7. - 2. i)\times  10^{-5}$$

\subsubsection {\bf  Hard scattering corrections}

The Hard scattering corrections and the Weak annihilation contributions are estimated at the scale $\mu = \sqrt{m_b\Lambda_h} = 1.45$. The Hard scattering contributions $H_i^-$ in (\ref{BB-1e}) are given by the expressions
$$H_i^- = - {18 f_{B_s} f_{V_2}^\perp(\mu_h) \over m_B m_b F_-^{B \to V_1}(0) } {m_b \over \lambda_B }\ (X_H-1)\ , \qquad \   i \in \{1, 2, 3, 4, 9, 10 \}$$
\beq
\label{B-13e}
\ \ \ \ \ H_i^- = {18 f_{B} f_{V_2}^\perp(\mu_h) \over m_{B_s} m_b F_-^{B \to V_1}(0) } {m_b \over \lambda_B }\ (X_H-1)\ , \qquad \qquad \qquad \ \ \ \  i \in \{5, 7 \}
\eeq
$$H_i^- = - {9 f_B f_{V_2} \over m_B m_b F_-^{B \to V_1}(0) } {m_b m_{V_1} \over m_{V_2}^2} {m_b\over \lambda_B }\ \ , \qquad \qquad \qquad  \qquad \ \  i \in \{6, 8 \}$$

\vskip 3 truemm

\noi Let us make a few remarks on these formulas.\par
\noi The quantities $H_i^-$ are inversely proportional to the form factor $F_-^{B \to V_1}(0)$. Notice that this gives a form-factor dependence of the coefficients $a_i^-$, the rest of the terms in (\ref{BB-1e}) being form factor independent.\par  

\noi The parameter $f_V^\perp$ is scale-dependent. For example, at $1\ {\rm GeV}$ one has \cite{BUCHALLA} $f_\phi^\perp(1{\rm GeV}) = 0.186\ {\rm GeV}$, and we need this parameter at the scale $\mu_h = 1.45\ {\rm GeV}$,  
\beq
\label{B-13e}
f_\phi^\perp(\mu_h) = f_\phi^\perp(1{\rm GeV}) \left({\alpha_s(\mu_h) \over \alpha_s (1{\rm GeV})} \right)^{C_F/\beta_0} = 0.179\ {\rm GeV}
\eeq
\noi Finally, we take for the parameter $\lambda_{B_s} = 0.2\ {\rm GeV}$ and $X_H$ parametrizes a divergent integral.
\beq
\label{C-9-1bise}
X_H = {\rm ln}\left( {m_{B_s} \over \Lambda_h}\right)
\eeq

\noi with $\Lambda_h = 0.5\ {\rm GeV}$.
Numerically one finds, for the example $V_2 = \phi$,
$$H_i^- = - 1.522\ , \qquad \qquad \qquad \   i \in \{1, 2, 3, 4, 9, 10 \}$$
\beq
\label{B-14e}
H_i^- = 1.488\ , \qquad \qquad \qquad \qquad \qquad \ \ \ \ i \in \{5, 7 \}
\eeq
$$H_i^- = - 2.816\ , \qquad \qquad \qquad  \qquad \qquad\ \  i \in \{6, 8 \}$$

\noi {\large \bf Summary on the coefficients $a_i^-$}

\noi Gathering the precedent results in (\ref{BB-1e}), and taking care of the two different scales $\mu = m_b$ and $\mu_h = \sqrt{\mu \Lambda_h}$ with $\Lambda_h = 0.5\ {\rm GeV}$ for the different terms, one finds the coefficients $a_i^-$ of Table~\ref{tab:14}.

\begin{table}[h]
\centering
\small

\begin{tabular}{|c|c|c|}
\hline 
Coefficient & ${\rm Re}(a_i^-)$ & ${\rm Im}(a_i^-)$\\ \hline \hline 
$a_1^-$ & $1.126$ & $0.029$ \\ \hline
$a_2^-$ & $-0.207$ & $-1.162$ \\ \hline
$a_3^-$ & $0.021$ & $0.005$ \\ \hline
$a_4^{-,u}$ & $-0.045$ & $-0.015$ \\ \hline
$a_4^{-,c}$ & $-0.043$ & $-0.001$ \\ \hline
$a_5^-$ & $-0.026$ & $-0.006$ \\ \hline
$a_7^{-,u}/\alpha$ & $1.052$ & $0.009$ \\ \hline
$a_7^{-,c}/\alpha$ & $1.024$ & $0.009$ \\ \hline
$a_9^{-,u}/\alpha$ & $-0.279$ & $-0.037$ \\ \hline
$a_9^{-,c}/\alpha$ & $-0.307$ & $-0.037$ \\ \hline
$a_{10}^{-,u}/\alpha$ & $0.292$ & $0.171$ \\ \hline
$a_{10}^{-,c}/\alpha$ & $0.293$ & $0.182$ \\ \hline
\end{tabular}

\caption{\label{tab:14} Central values of the coefficients $a_i^-$.}

\end{table}

\noi The values in the table are central values but one must take into account that these coefficients vary from one mode to another. We take into account the dispersion of these coefficients due to the different modes giving a central value and an error in Table~\ref{tab:6}.

\section {\Large \bf  The annihilation coefficients $b_i^h$}

The coefficients $b_i^h (h = 0, -)$ are given by in terms of building blocks $A_i^{ih}, A_i^{fh}$,
$$b_3^h = {C_F \over N_c^2} \left[C_3 A_1^{ih} + C_5 (A_3^{ih} + A_3^{fh}) +N_c C_6 A_3^{fh} \right] \ \ \ \ $$
\beq
\label{B-15e}
b_4^h = {C_F \over N_c^2} \left[C_4 A_1^{ih} + C_6 A_2^{ih} \right] \qquad \qquad \qquad \qquad \ \ \ \ \ 
\eeq
$$b_3^{EWh} = {C_F \over N_c^2} \left[C_9 A_1^{ih} + C_7 (A_3^{ih} + A_3^{fh}) +N_c C_8 A_3^{fh} \right]$$
$$b_4^{EWh} = {C_F \over N_c^2} \left[C_{10} A_1^{ih} + C_8 A_2^{ih} \right] \qquad \qquad \qquad \qquad $$

\subsection {\large \bf  Longitudinal $h = 0$ building blocks $A_i^{i0}, A_i^{f0}$}

For $h = 0$ the building blocks $A_i^{i0}, A_i^{f0}$ are given by the expressions \cite{BUCHALLA}
\beq
\label{C-1e}
A_1^{i0} = A_2^{i0} \ , \qquad \qquad A_3^{i0} = 0
\eeq

\noi and 
$$A_1^{i0} \simeq 18 \pi \alpha_s \left[\left(X_A - 4 + {\pi^2 \over 3} \right) + (r_\perp^V)^2 (X_A-2)^2\right]$$
\beq
\label{C-2e}
A_3^{f0} \simeq - 36 \pi \alpha_s r_\perp^V (2X_A^2-5X_A + 2)
\eeq

\noi With the parametrization
\beq
\label{C-3e}
X_A = \int_0^1 {dx \over x} = \left (1 + \rho_A e^{i \phi_A} \right) \ln {m_B \over \Lambda_h}
\eeq

\noi and the values $\rho_A = 0.6, \phi_A = - 40^0$ of ref. \cite{BENEKE-2007}, one finds the annihilation coefficients for the longitudinal amplitude of Table~\ref{tab:15} below. One finds a very large coefficient $b_3^0$, comparable in absolute magnitude to the transverse one $b_3^-$.

\begin{table}[h]
\centering
\small

\begin{tabular}{|c|c|c|}
\hline 
Coefficient & ${\rm Re}(b_i^0)$ & ${\rm Im}(b_i^0)$\\ \hline \hline 
$b_1^0$ & $9.687$ & $-4.053$ \\ \hline
$b_2^0$ & $-3.045$ & $1.274$ \\ \hline
$b_3^0$ & $3.407$ & $-3.811$ \\ \hline
$b_4^0$ & $-1.204$ & $0.504$ \\ \hline
$b_3^{0EW}$ & $-0.124$ & $0.081$ \\ \hline
$b_4^{0EW}$ & $0.035$ & $-0.015$ \\ \hline
\end{tabular}

\caption{\label{tab:15} Central values of the coefficients $b_i^0$.}

\end{table}

\vskip 5 truemm

\noi {\large \bf Summary on the coefficients $b_i^0$}

\noi The values in the table are central values but one must take into account that these coefficients vary from one mode to another due to the dependence on the form factors. We take into account the dispersion of these coefficients due to the different modes giving a central value and an error in Table~\ref{tab:7}.

\subsection {\large \bf  Transverse $h = -$ building blocks $A_i^{i-}, A_i^{f-}$}

For $h = -$ the building blocks $A_i^{i0}, A_i^{f0}$ are given by the expressions (69) of \cite{BENEKE-2007}

$$A_1^{i-} = A_2^{i-} = 18 \pi \alpha_s {m_{V_1}m_{V_2} \over m_{B}^2} \left({1 \over 2} X_L + {5 \over 2} - {\pi^2 \over 3} \right)$$
\beq
\label{B-16e}
A_3^{i-} = 18 \pi \alpha_s \left({m_{V_1} \over m_{V_2}} r_\perp^{V_2} - {m_{V_2} \over m_{V_1}} r_\perp^{V_1} \right) (X_A^2-2X_A+2) \ \ \ 
\eeq
$$A_3^{f-} = 36 \pi \alpha_s \left({m_{V_1} \over m_{V_2}} r_\perp^{V_2} + {m_{V_2} \over m_{V_1}} r_\perp^{V_1} \right)
 (2 X_A^2-5X_A+3) \ \ \ $$

\noi where the running of $r_\perp^{V_2}(\mu)$ is given by \cite{BUCHALLA} 
\beq
\label{B-17e}
r_\perp ^{V_2}(\mu) = {2m_{V_2} f_{V_2}^\perp(\mu) \over m_b(\mu) f_{V_2}} = {2m_{V_2} f_{V_2}^\perp(1 {\rm GeV}) \over m_b(m_b) f_{V_2}} \left[ {\alpha_s(\mu) \over \alpha_s(m_b)} \right]^{-3C_F/\beta_0} \left[ {\alpha_s(\mu) \over \alpha_s(1 {\rm GeV})} \right]^{C_F/\beta_0}
\eeq

\noi and must be computed at the scale $\mu_h$.

\vskip 3 truemm
\noi For the example $f_\phi^\perp(1 {\rm GeV}) = 0.186\ {\rm GeV}$ \cite{BUCHALLA} or $f_\phi^\perp(2 {\rm GeV}) = 0.175\ {\rm GeV}$ \cite{BENEKE-2007}, one finds consistently, for $\mu = \sqrt{m_b\Lambda_h} = 1.45\ {\rm GeV}$
\beq
\label{B-18e}
r_\perp ^\phi(1.45\ {\rm GeV}) = 0.318
\eeq 

\noi Finally, $X_A$ and $X_L$ parametrize divergent integrals, as given by \cite{BENEKE-2007},
\beq
\label{C-9-1e}
X_A = \left(1+0.6\ e ^{-i 40^0} \right) {\rm ln}\left( {m_{B_s} \over \Lambda_h}\right) \ , \qquad X_L = {m_b \over \Lambda_{QCD}} 
\eeq

\noi Numerically one finds, for the example $\overline{B}_s \to \phi \phi$,
\beq
\label{B-19e}
A_1^{i-} = A_2^{i-} = 5.888 \ , \qquad A_3^{i-} = 0 \ , \qquad A_3^{f-} = 97.450 - 98.673 i
\eeq

\noi and one gets the coefficients given in Table~\ref{tab:16}. One sees that the annihilation is largely dominated by $b_3^-$.\par

\begin{table}[h]
\centering
\small
\begin{tabular}{|c|c|c|}
\hline 
Coefficient & ${\rm Re}(b_i^-)$ & ${\rm Im}(b_i^-)$\\ \hline \hline 
$b_1^-$ & $0.691$ & $0.$ \\ \hline
$b_2^-$ & $-0.217$ & $0.$ \\ \hline
$b_3^-$ & $-3.764$ & $3.848$ \\ \hline
$b_4^-$ & $-0.086$ & $0.$ \\ \hline
$b_3^{-EW}$ & $0.031$ & $-0.040$ \\ \hline
$b_4^{-EW}$ & $0.003$ & $0.$ \\ \hline
\end{tabular}

\caption{\label{tab:16}  Central values of the coefficients $b_i^-$ for the example $\overline{B}_s \to \phi \phi$.}

\end{table}

\vskip 5 truemm

\noi {\large \bf Summary on the coefficients $b_i^-$}

\noi The values in the table are central values in a particular case but one must take into account that these coefficients vary from one mode to another. We take into account the dispersion of these different coefficients due to the different modes giving a central value and an error in Table~\ref{tab:7}.

\vskip 10 truemm

\noi {\Large \bf Acknowledgements}

\vskip 5 truemm

We are very much indebted to Martin Beneke and Gerhard Buchalla for providing us useful and detailed information on the QCD Factorization scheme applied to the decays of $B$ mesons into two light vectors mesons.


\begin{thebibliography}{99} 

\bibitem{PDG-2022} R. L. Workman et al. (Particle Data Group), Progr. Theor. Exp. Phys. 2022, 083C01 (2022) and 2023 update.

\bibitem{KAGAN} A. Kagan, {\it Polarization in $B \to VV$ decays}, Phys. Lett. B {\bf 601} (2004) 151, arXiv:hep-ph/0405134.

\bibitem{BENEKE-2007} M. Beneke, J. Rohrer and D. Yang,  {\it Branching fractions, polarisation and asymmetries of $B \to VV$ decays}, Nucl. Phys. B {\bf 774} (2007) 64, hep-ph/0612290.

\bibitem{CHENG-YANG} H.-Y. Cheng and K.-C. Yang, {\it Branching ratios and polarization in $B \to VV, VA, AA$ decays}, Phys. Rev. D {\bf 78} (2008) 094001, Phys. Rev. D {\bf 79} (2009) 039903 (erratum), arXiv:0805.0329.

\bibitem{BUCHALLA} M. Bartsch, G. Buchalla and C. Kraus, {\it $B \to V_L V_L$ Decays at NLO in QCD}, arXiv:0810.0249.

\bibitem{BBNS-01} M. Beneke, G. Buchalla, M. Neubert and C. Sachrajda, {\it QCD factorization in $B \to \pi K, \pi \pi$ decays and extraction of Wolfenstein parameters}, Nucl. Phys. B {\bf 606} (2001) 245, arXiv:hep-ph/0104110.

\bibitem{BBNS-99} M. Beneke, G. Buchalla, M. Neubert and C.T. Sachrajda, {\it QCD factorization for $B \to \pi \pi$ decays: Strong phases and CP violation in the heavy quark limit}, Phys. Rev. Lett. {\bf 83} (1999) 1914, arXiv:hep-ph/9905312.

\bibitem{BBNS-00} M. Beneke, G. Buchalla, M. Neubert and C.T. Sachrajda, {\it QCD factorization for exclusive, nonleptonic $B$ meson decays: General arguments and the case of heavy light final states}, Nucl. Phys. B {\bf 591} (2000) 313, arXiv:hep-ph/0006124.

\bibitem{BENEKE-NEUBERT} M. Beneke and M. Neubert, Nucl. Phys. B {\bf 675}, (2003) 333, arXiv:hep-ph/0308039.

%\bibitem{NIERSTE-2000} I. Dunietz, R. Fleischer and U. Nierste, {\it In Pursuit of New Physics with $B_s$ Decays}, Phys. Rev. D {\bf 63} (2001) 114015, arXiv:hep-ph/0012219.

%\bibitem{NIERSTE-2009} U. Nierste, {\it Three Lectures in Meson Mixing and CKM phenomenology}, Contribution to Helmholz International Summer School on Heavy Quark Physics, arXiv:0904.1869.

%\bibitem{BUCHALLA-BURAS-1990} G. Buchalla, A. Buras and M. Harlander, {\it The Anatomy of $\epsilon' / \epsilon$ in the Standard Model}, Nucl. Phys. B {\bf 337} (1990) 313.

%\bibitem{CKMfitter} The CKMfitter Group, J. Charles et al., {\it Current status of the Standard Model CKM fit and constraints on $\Delta F = 2$ New Physics}, Phys. Rev. D {\bf 91} (2015) 7, 073007, arXiv:1501.05013 (2015).

%\bibitem{pdg:1} P.A. Zyla et al. (Particle Data Group), Prog. Theor. Exp. Phys. 2020, 083C01 (2020).

%\bibitem{fccee:1}M. Bicer, et al., First look at the physics case of TLEP, J. High Energy Phys. 01 (2014) 164, https://doi.org /10.1007/JHEP01(2014)164, arXiv:1308.6176.

%\bibitem{fccee:2} A. Abada, et al., FCC Collaboration, Eur. Phys. J. C 79(6) (2019) 474, https://doi.org /10.1140 /epjc/s10052-019-6904-3.

%\bibitem{fccee:3} A. Abada, et al., FCC Collaboration, Eur. Phys. J. ST 228(2) (2019) 261, https://doi.org/10.1140/epjst/e2019-900045-4.

%\bibitem{ALEKSAN} R. Aleksan, L. Oliver and E. Perez, Phys. Rev. D {\bf 105} (2022) 5, 053008, arXiv:2107.02002 [hep-ph].

%\bibitem{BALL-ZWICKY-2} P. Ball and R. Zwicky, {\it $B_{d,s} \to \rho, \omega, K^*, \phi$ decay form factors from light-cone sum rules revisited},    Phys. Rev. D {\bf 71} (2005) 014029, arXiv:hep-ph/0412079.

%\bibitem{BALL-BRAUN} P. Ball and V. Braun, {\it Exclusive Semileptonic and Rare B-Meson Decays in QCD}, Phys. Rev. D {\bf 58} (1998) 094016, arXiv:hep-ph/9805422.

\bibitem{AO:1} R. Aleksan and L. Oliver, {\it $\rm B_{d,s}\to K^{*0}\overline{K}^{*0}$ decays, a serious problem for the Standard Model}, arXiv:2312.07198[hep-ph].
\bibitem{BUCHALLA-HARLANDER-1995} G. Buchalla, A. Buras and E. Lautenbacher, {\it Weak decays beyond leading logarithms}, Rev. Mod. Phys. {\bf 68} (1996) 1125, arXiv:hep-ph/9512380.

\end{thebibliography}
\enddocument